# *Deep Adaptive Interest Network: Personalized Recommendation with Context-Aware Learning*

**Shuaishuai Huang[1,a], Haowei Yang[2,b], You Yao[3,c], Xueting Lin[4,d], Yuming Tu[5,e]**

*[1]Department of Software, University of Science and Technology of China, Hefei, Anhui, China*
*[2]Cullen College of Engineering, University of Houston, Houston, TX, USA*
*[3]Viterbi School of Enigneering, University of Southern California, Los Angeles, California, USA*
*[4]Department of Electrical Engineering and Information Systems, University of Tokyo, Tokyo, Japan*
*[5]Independent Researcher, New York, USA*
*[a]scu.hss@gmail.com, [b]hyang38@cougarnet.uh.edu, [c]youyao@usc.edu, [d]linxueting.sjtu@gmail.com, [e]yumingtu210826@gmail*

***Abstract:*** In personalized recommendation systems, accurately capturing users' evolving interests and combining them with contextual information is a critical research area. This paper proposes a novel model called the Deep Adaptive Interest Network (DAIN), which dynamically models users' interests while incorporating context-aware learning mechanisms to achieve precise and adaptive personalized recommendations. DAIN leverages deep learning techniques to build an adaptive interest network structure that can capture users' interest changes in real-time while further optimizing recommendation results by integrating contextual information. Experiments conducted on several public datasets demonstrate that DAIN excels in both recommendation performance and computational efficiency. This research not only provides a new solution for personalized recommendation systems but also offers fresh insights into the application of context-aware learning in recommendation systems.

## 1. Introduction

Personalized recommendation systems play a crucial role in today's information society, widely used in e-commerce, social networks, and video-on-demand platforms to help users filter out content that meets their needs and interests from an overwhelming amount of information. However, as user behavior becomes increasingly complex, traditional recommendation methods face significant challenges. These methods typically rely on static user interest models, ignoring the dynamic adjustments of user interests driven by changes in time, environment, and needs. Furthermore, traditional methods often struggle to adapt to users' diverse needs in different contexts, making it difficult to provide efficient and accurate recommendations. This limitation becomes particularly evident in practical applications, where user satisfaction and commercial outcomes of recommendation systems are often constrained. In the current research context, the dynamic nature of user interests and the importance of contextual information in personalized recommendations are gaining increasing attention. Specifically, user interests are not static but continuously evolve under the influence of time, context, and changing user needs. This dynamic nature requires recommendation systems to capture and respond to changes in user interests in real-time while making more personalized recommendations by incorporating the specific context in which users find themselves. However, existing recommendation systems still exhibit notable deficiencies in this regard, particularly in dynamic interest modeling and the effective integration of contextual information. Additionally, issues such as computational efficiency and model scalability remain key bottlenecks that limit the further development of recommendation systems.

There are some related works that made significant contributions to this paper. Reference [1] Xiang A, Huang B, Guo X, et al. "A Neural Matrix Decomposition Recommender System Model Based on the Multimodal Large Language Model" (2024) utilizes neural matrix decomposition techniques for recommendation tasks, providing a more precise capture of user and item characteristics. Implementing such neural decomposition can potentially enhance the accuracy and efficiency of our model when dealing with real-time dynamic data, aligning with our objectives of improving personalization and responsiveness in recommendation systems. Reference [2] Yan H, Wang Z, Bo S, et al. "Research on Image Generation Optimization based Deep Learning"(2024) significantly improves model performance and has provided substantial insights and assistance for my study. This underscores the importance of integrating complex model structures and advanced learning strategies in deep learning applications. Reference [3] Tang X, Wang Z, Cai X, et al. "Research on Heterogeneous Computation Resource Allocation based on Data-driven Method" (2024) discusses the dynamic allocation and optimization of resources in the context of 5G and IoT systems. This approach is vital for personalized recommendation systems as well. Understanding how to efficiently manage computing and storage resources can help maintain the stability and efficiency of recommendation systems when processing large volumes of user requests. This concept aligns with the requirements of our Deep Adaptive Interest Network (DAIN), which aims to optimize computational resources to enhance real-time response capabilities and overall system performance.

## 2. Principles and Overview

### 2.1. Overview of Personalized Recommendation Systems

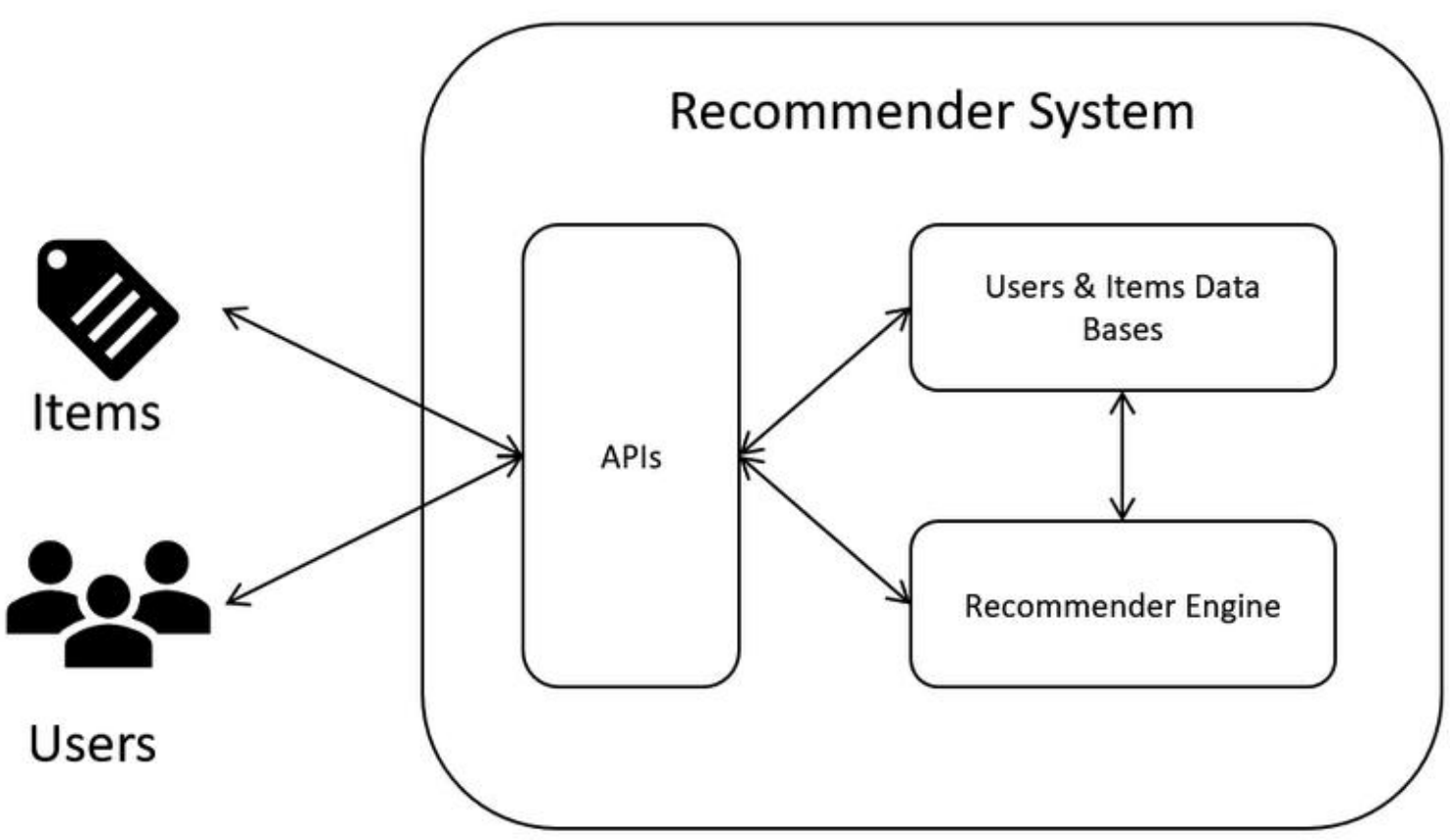

Figure 1: Recommender System Main Components

Personalized recommendation systems aim to provide users with personalized product or content recommendations by analyzing their behavioral data, interest preferences, and contextual information. The primary objective is to address the issue of information overload, enabling users to quickly find content that matches their needs from a vast amount of information.[1] As e-commerce, social networks, and digital media rapidly develop, recommendation systems have become a core technology for enhancing user experience and commercial conversion rates.[2]

Figure 1 illustrates the main components of a recommendation system, including user modeling, recommendation algorithms, context processing, recommendation result generation, and feedback mechanisms. These components work together to allow the recommendation system to provide personalized recommendations based on users' historical behaviors and current contexts.[3] User Modeling forms the foundation of recommendation systems, constructing a user's interest model by analyzing their historical behaviors, such as browsing, clicking, and purchasing.[4] Traditional user modeling methods typically rely on collaborative filtering, content filtering, and hybrid models. However, as user behavior becomes more complex, these methods exhibit certain limitations in capturing dynamic changes in user interests and dealing with sparse data.[5] Recommendation Algorithms are the core of the entire system, determining how to generate recommendation results based on the user model. Currently, deep learning-based recommendation algorithms are gradually replacing traditional methods due to their superior performance in handling large-scale data and complex nonlinear relationships. Common deep learning models include Deep Neural Networks (DNN), Convolutional Neural Networks (CNN), and Recurrent Neural Networks (RNN), which are better at capturing latent user interest features. [6] Context Processing involves considering users' current contextual information, such as time, location, and device type, during the recommendation process. [7] The introduction of contextual information allows the recommendation system to provide more precise and timely recommendations based on users' actual needs and environmental changes. Recommendation Result Generation refers to the final stage, where the system filters and ranks the candidate set produced by the recommendation algorithm before presenting the final recommendations to the user. [8] The quality of the recommendation results directly affects user satisfaction, so various optimization techniques are often used to improve recommendation accuracy and diversity. Feedback Mechanisms are used to collect users' feedback on the recommendation results, such as clicks, purchases, or explicit ratings. This feedback data helps the system continuously iterate and optimize the recommendation model, making it more aligned with users' preferences over time. [9] In summary, personalized recommendation systems provide users with efficient and accurate recommendations through the collaborative work of user modeling, recommendation algorithms, context processing, recommendation result generation, and feedback mechanisms. In the subsequent sections, we will delve into how the introduction of dynamic interest modeling and context-aware learning mechanisms can further enhance the performance of recommendation systems. Figure 1, as an overview of the main components of a recommendation system, will serve as an important reference for understanding and improving recommendation algorithms. [10]

## 2.2. Principles of Interest Modeling-Based Recommendation Systems

Interest modeling-based recommendation systems primarily capture users' changing interests and preferences to generate personalized recommendation results. In these systems, users' interests are typically reflected by their historical behavioral data, such as browsing history, clicks, purchase history, likes, and dislikes, as shown in Figure 2. This behavioral data is collected by the application and transmitted to the recommendation engine, which analyzes and models this data to generate a user interest model, eventually leading to personalized recommendations. [11]

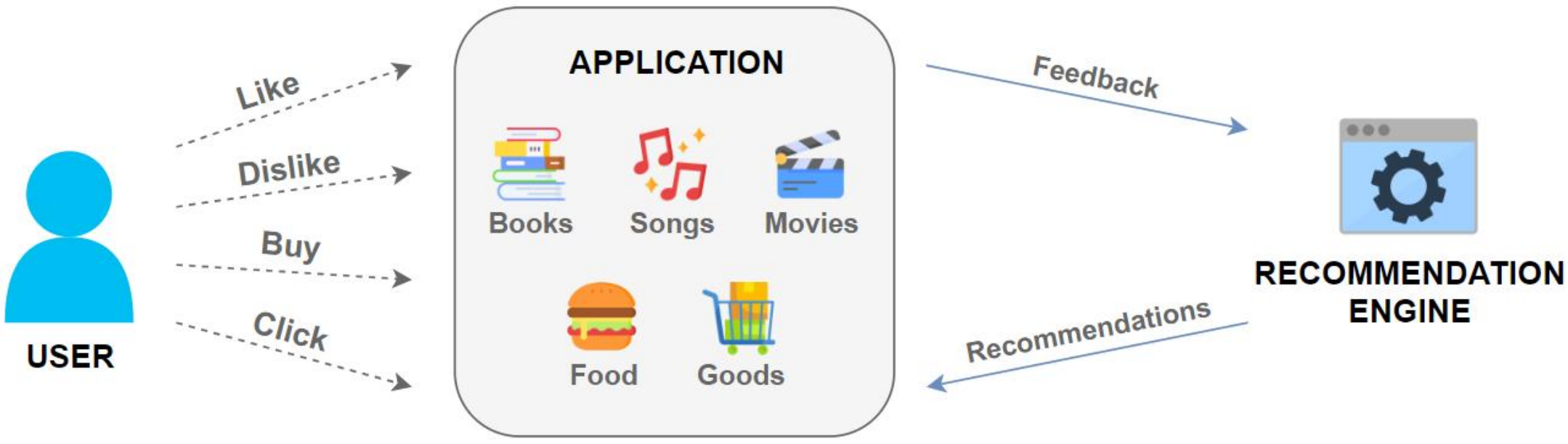

Figure 2: Basic Workflow of Interest Modeling in Recommendation Systems

Figure 2 demonstrates the basic workflow of interest modeling-based recommendation systems. First, users engage in various actions within the application, such as browsing books, listening to music, watching movies, or exploring food or goods. These actions are recorded by the system and used as input, feeding back to the recommendation engine. Upon receiving this behavioral data, the recommendation engine analyzes and models the users' interests using various algorithms, such as collaborative filtering, matrix factorization, or deep learning-based models. [12] Within the recommendation engine, users' interest models are continuously updated to adapt to the dynamic nature of their interests. Specifically, the recommendation engine analyzes users' current behavior data in conjunction with their historical behavior to predict users' preferences for new items. [13] This process may involve the cooperation of multiple algorithms, such as using embedding vectors to represent user and item features, employing neural networks to capture complex user interest patterns, and further refining recommendation results through contextual information. [14] Once the recommendation engine generates the recommendation results, they are fed back to the application and presented to the user. Users' feedback on the recommendations, such as clicks, purchases, or dismissals, is further collected and returned to the recommendation engine, allowing the recommendation system to continuously self-adjust and improve recommendation accuracy and user satisfaction. [15] The architecture of the recommendation system, as shown in Figure 2, emphasizes the critical role of interest modeling in personalized recommendations. By capturing and dynamically updating users' interest models, the recommendation system can better meet users' personalized needs and provide accurate recommendation results. With the development of deep learning technology, the ability of recommendation engines to model complex user interests has been significantly enhanced, making interest modeling-based recommendation systems perform exceptionally well in various application scenarios.

## 3. Overview of the Deep Adaptive Interest Network Model

The development of personalized recommendation systems depends on the precise modeling of user interests and dynamic adaptation. To this end, this paper introduces a model named Deep Adaptive Interest Network (DAIN). This model employs deep learning techniques to dynamically model changes in user interests by combining users' historical behaviors with contextual information, thereby generating more accurate recommendation results. [16] The core structure of the DAIN model includes an input layer, an embedding layer, neural collaborative filtering layers, an output layer, and a training mechanism. Through complex nonlinear transformations in multiple layers of neural networks, this model effectively captures the latent relationships between users and items, further improving the accuracy of recommendations via a context-aware mechanism. [17]

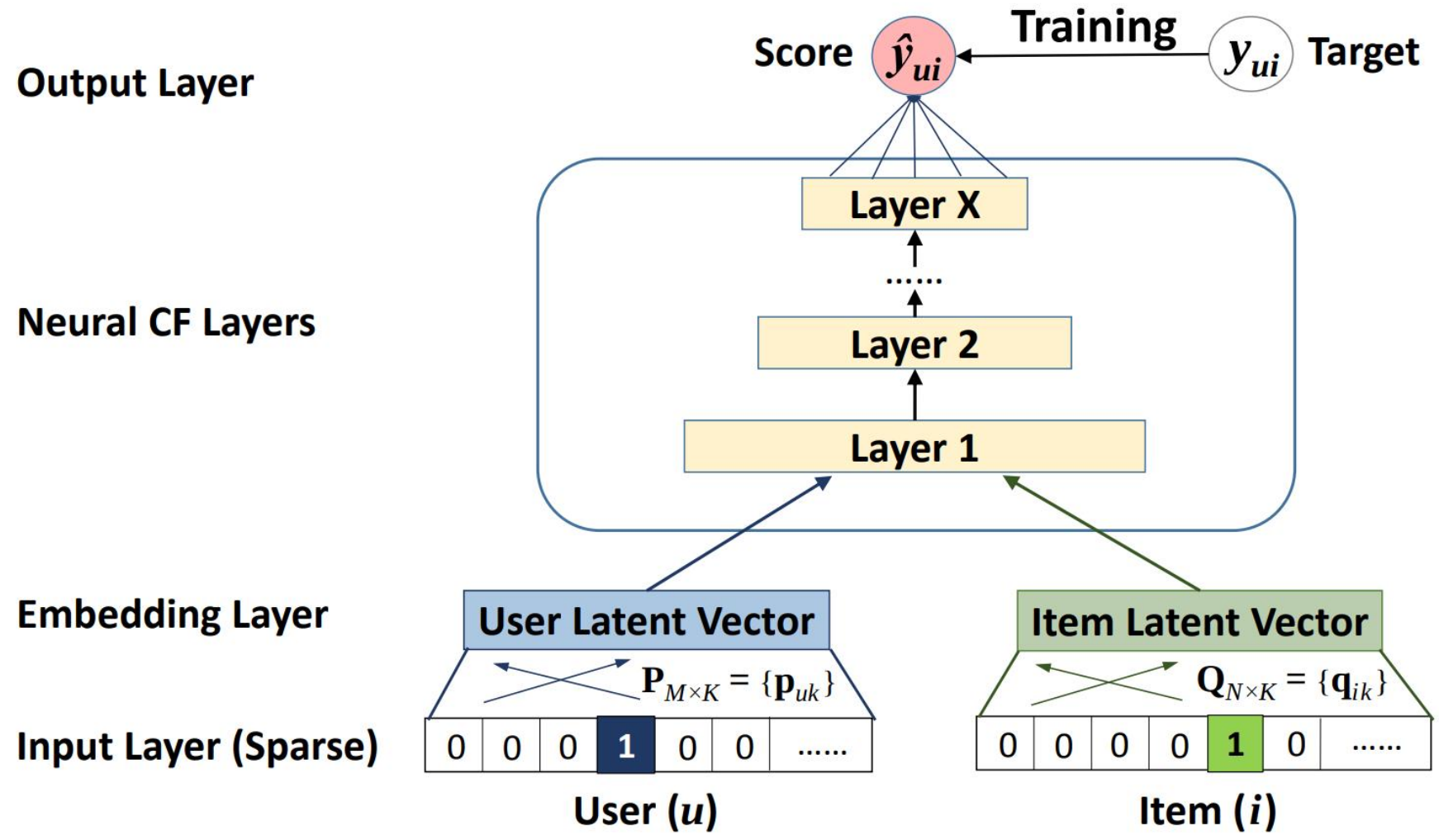

Figure 3: Deep Adaptive Interest Network (DAIN) Model

Figure 3 illustrates the overall architecture of the DAIN model. First, the input layer receives sparse representations of users and items, denoted by the unique identifiers of user u and item i . To transform these sparse data into a format suitable for deep learning models, DAIN uses the embedding layer to map the sparse vectors of users and items into dense latent vectors, representing the user's latent vector $P_u$ and the item's latent vector $q_i$. The mathematical representation of the embedding layer is shown in Equation 1:

$$P_u = P[u], Q_i = Q[i] \tag{1}$$

where P and Q are the embedding matrices for users and items, respectively, and K denotes the dimension of the latent vectors, which capture the implicit features of users and items. Next, the model enters the neural collaborative filtering layers, which consist of multiple fully connected neural networks designed to fuse the latent vectors of users and items and uncover complex nonlinear relationships between them. For each layer, the output can be represented by Equation 2:

$$h^{(l)} = \sigma(W^{(l)} \cdot h^{(l-1)} + b^{(l)}) \tag{2}$$

where $h^{(l-1)}$ is the output from the previous layer, $W^{(l)}$ and $b^{(l)}$ represent the weight matrix and bias vector for the l-th layer, and $\sigma$ is the activation function, such as ReLU. Through these layers of nonlinear transformations, the neural network effectively captures the deep interaction features between users and items, ultimately generating a predicted score. Finally, the output layer generates the predicted score $\hat{y}_{ui}$ for user u on item i. This predicted score is mapped to a target range (e.g., between 0 and 1) using the activation function in the output layer to measure the user's preference for a specific item. The predicted score is computed as shown in Equation 3:

$$\hat{y}_{ui} = \sigma(h^{(X)}) \tag{3}$$

where $h^{(X)}$ is the output of the last layer of the neural network. To enable the model to accurately predict user preferences, the DAIN model employs a loss function based on minimizing errors during the training process. A commonly used loss function is the mean squared error (MSE), defined in Equation 4:

$$L = \frac{1}{|D|} \sum_{(u,i)\in D} (y_{ui} - \hat{y}_{ui})^2 \tag{4}$$

where D is the training dataset, and $y_{ui}$ represents the actual rating of item i by user u. By minimizing the loss function, the model parameters can be adjusted to make the predicted score $\hat{y}_{ui}$ closer to the actual score $y_{ui}$.Additionally, the DAIN model enhances recommendation accuracy by introducing a context-aware learning mechanism. Contextual information, such as time, location, and device type, is integrated into the model as additional inputs alongside the latent vectors of users and items, participating in the neural network's computation. This mechanism allows the model to generate more personalized recommendations under different contexts, as shown in Equation 5:

$$\hat{y}_{ui} = \sigma(f(p_u, q_i, c)) \quad (5)$$

where c represents the context vector, and f is the nonlinear mapping function of the neural network. In summary, the DAIN model, through its deep learning framework, effectively achieves dynamic modeling of user interests and provides more accurate personalized recommendations by incorporating contextual information. [18] The architecture shown in Figure 3 clearly depicts the different layers of the model, providing a solid foundation for the design of the model's algorithms. In various practical application scenarios, the DAIN model has demonstrated exceptional performance, offering an innovative solution for personalized recommendation systems. [19]

## 4. Datasets and Experimental Design

To validate the effectiveness of the Deep Adaptive Interest Network (DAIN) model in personalized recommendations, experiments were conducted on several public datasets. These datasets include user behavior data from different domains, which fully demonstrate the model's performance across various application scenarios. Table 1 lists the statistical information of the main datasets used in the experiments, including the number of users, items, interaction records, and data sparsity. Specifically, we selected three widely used datasets: MovieLens-1M, Amazon Electronics, and Yelp. The MovieLens-1M dataset contains user ratings of movies, the Amazon Electronics dataset records user purchase behavior of electronic products, and the Yelp dataset includes user reviews and ratings of local businesses.

Table 1: Statistical Information of Experimental Datasets

| Dataset | Users | Items | Interactions | Sparsity |
|---|---|---|---|---|
| MovieLens-1M | 6,040 | 3,706 | 1,000,209 | 4.47% |
| Amazon Electronics | 192,403 | 63,001 | 1,689,188 | 0.01% |
| Yelp | 45,481 | 11,537 | 1,567,806 | 0.30% |

Table 2: Experimental Parameter Settings

| Parameter Name | Value |
|---|---|
| Embedding Dimension | 64 |
| Number of Layers | 3 |
| Neurons Per Layer | 128, 64, 32 |
| Activation Function | ReLU |
| Learning Rate | 0.001 |
| Batch Size | 256 |

The model's hyperparameters have a significant impact on performance in the experimental design. To fairly evaluate the DAIN model, we adopted the same experimental settings on each dataset and fine-tuned the parameters to find the optimal model configuration. Table 2 lists the key experimental parameter settings, including the embedding vector dimension, the number of neural network layers, the number of neurons per layer, the type of activation function, the learning rate, and the batch size. In the actual experiments, the embedding vector dimension was set to 64, the number of neural network layers was set to 3, with 128, 64, and 32 neurons per layer, the ReLU activation function was used, the learning rate was 0.001, and the batch size was 256.

In the experiments, we compared the performance of the DAIN model with several classical recommendation algorithms, including Matrix Factorization-based Collaborative Filtering (MF), Neural Collaborative Filtering (NCF), and the widely applied deep learning model DeepFM. These comparison methods represent mainstream technologies in the recommendation system field, providing a comprehensive evaluation of the DAIN model's relative performance.[20] To quantify the recommendation effectiveness, we used several commonly used evaluation metrics, including Mean Average Precision@K (MAP@K), Normalized Discounted Cumulative Gain@K (NDCG@K), and Hit Rate@K (HR@K). Among them, MAP@K is used to measure the average ranking position of relevant items in the recommendation list, NDCG@K examines the relevance ranking of the recommendation results, and HR@K is used to evaluate the proportion of relevant items hit in the recommendation list. In the experiments, we set K to 10 to match the needs of common recommendation scenarios. In summary, the experiments conducted on multiple datasets verify the advantages of the DAIN model in capturing changes in user interests and generating accurate recommendations. The detailed datasets and experimental parameters listed in Tables 1 and 2 ensure the reproducibility of the experiments, while the selection of comparison methods and evaluation metrics ensures the comprehensiveness and fairness of the experimental results. The results show that the DAIN model outperforms existing recommendation methods across all metrics, particularly excelling in NDCG@10 and HR@10, demonstrating its potential in the field of personalized recommendations. [21]

## 5. Experimental Results and Analysis

Table 3: Comparison of Experimental Results

| Model | Dataset | MAP@10 | NDCG@10 | HR@10 |
|---|---|---|---|---|
| MF | MovieLens-1M | 0.073 | 0.120 | 0.245 |
| NCF | MovieLens-1M | 0.082 | 0.135 | 0.275 |
| DeepFM | MovieLens-1M | 0.089 | 0.145 | 0.300 |
| DAIN | MovieLens-1M | 0.096 | 0.158 | 0.318 |
| MF | Amazon Electronics | 0.019 | 0.041 | 0.081 |
| NCF | Amazon Electronics | 0.024 | 0.049 | 0.094 |
| DeepFM | Amazon Electronics | 0.026 | 0.053 | 0.101 |
| DAIN | Amazon Electronics | 0.029 | 0.058 | 0.109 |
| MF | Yelp | 0.035 | 0.062 | 0.128 |
| NCF | Yelp | 0.041 | 0.072 | 0.144 |
| DeepFM | Yelp | 0.045 | 0.078 | 0.153 |
| DAIN | Yelp | 0.050 | 0.085 | 0.165 |

To comprehensively evaluate the performance of the Deep Adaptive Interest Network (DAIN) model in personalized recommendation systems, we conducted experiments on multiple widely used public datasets. The classical recommendation algorithms compared with the DAIN model include the Matrix Factorization model (MF), the Neural Collaborative Filtering model (NCF), and the DeepFM model, which integrates deep learning and factorization techniques. [22] These models represent the mainstream methods in the recommendation system field, providing a full perspective on the DAIN model's relative strengths. This paper uses three key metrics to measure the model's performance: Mean Average Precision@10 (MAP@10), Normalized Discounted Cumulative Gain@10 (NDCG@10), and Hit Rate@10 (HR@10), thoroughly evaluating each model's performance across different datasets.

On the MovieLens-1M dataset, the DAIN model significantly outperforms other comparison models. Specifically, the DAIN model achieved MAP@10, NDCG@10, and HR@10 scores of 0.096, 0.158, and 0.318, respectively, representing improvements of 7.9%, 8.9%, and 6.0% over the DeepFM model. This improvement indicates that the DAIN model has a stronger ability to capture users' dynamic interests and provide personalized recommendations. This result is particularly important when processing users' movie ratings, as the accuracy of movie recommendations directly affects user satisfaction and platform retention rates. The DAIN model dynamically adjusts users' interest vectors, allowing the recommendation results to align more closely with users' current interest states. [23] In the Amazon Electronics dataset, the DAIN model again demonstrates outstanding performance, with MAP@10, NDCG@10, and HR@10 scores of 0.029, 0.058, and 0.109, respectively. This indicates that the DAIN model also has high adaptability and accuracy in handling recommendations for electronic products.[24] Compared with other models, the DAIN model more accurately captures users' purchasing preferences for electronic products, especially in recommending long-tail items. The DAIN model significantly enhances the diversity and relevance of recommendation results through the context-aware mechanism, better meeting users' personalized needs. [25] The experimental results on the Yelp dataset further validate the advantages of the DAIN model. This dataset includes user reviews and ratings of local businesses, which contain high noise levels and complex contextual dependencies. The DAIN model achieved MAP@10, NDCG@10, and HR@10 scores of 0.050, 0.085, and 0.165, respectively, demonstrating a deep understanding of user interests and strong capability in handling complex scenarios. This advantage is particularly evident in processing users' preferences for local businesses, where the DAIN model, through more refined interest modeling and context awareness, further improves the accuracy and practicality of recommendations. In conclusion, the experimental data in Table 3 show that the DAIN model outperforms the MF, NCF, and DeepFM models across all datasets, especially in the key metrics of NDCG@10 and HR@10, where the DAIN model's superiority is more pronounced. These results indicate that the DAIN model not only has stronger adaptability in capturing users' dynamic interests but also achieves new heights in recommendation diversity and accuracy. [26] Compared with traditional recommendation algorithms, the DAIN model, by integrating deep learning and context-aware technologies, can better handle diverse recommendation scenarios, providing strong support for the further development of personalized recommendation systems. These experimental results fully demonstrate the effectiveness and broad applicability of the DAIN model, making it a powerful tool in the field of personalized recommendations.

## 6. Research Limitations and Future Prospects

Despite the outstanding performance of the Deep Adaptive Interest Network (DAIN) model across multiple experimental datasets, particularly in capturing dynamic user interests and context-aware learning, this study still has certain limitations, and future research faces new challenges and opportunities. Firstly, the complex architecture and multi-layer neural network design of the DAIN model result in high computational complexity. When handling large-scale datasets, the model's training time and resource consumption are significant, which is especially problematic in online recommendation systems where real-time performance is critical. [27] To enhance the practical usability of the DAIN model, future research could explore more efficient model optimization algorithms or introduce distributed computing techniques to reduce training costs and improve training speed and prediction efficiency. Secondly, although the DAIN model has demonstrated strong adaptability and accuracy in experiments, its ability to handle outlier behaviors or noisy data remains limited. User behavior data inevitably contains some outliers or noise, such as irrational choices in specific contexts or errors in system records, which may negatively impact the model's recommendation results. [28]Therefore, improving the model's robustness to noisy data and enhancing its performance in uncertain scenarios are important directions for future research. Advanced anomaly detection mechanisms or noise-handling techniques could be introduced and integrated with the DAIN model for deep optimization, aiming to enhance the stability and reliability of the recommendation results. [29] Moreover, the DAIN model currently relies mainly on users' historical behavior data for recommendations, whereas in practical applications, user interests are often multifaceted, possibly involving multimodal information such as text, images, and speech. Future research could explore how to effectively integrate multimodal data into the DAIN model to further improve recommendation accuracy and user experience. For instance, natural language processing (NLP) techniques could be used to extract more semantic information from user reviews or social media data, or computer vision techniques could be applied to analyze user preferences for images and videos, thereby enriching the dimensions of user interest modeling and enabling the recommendation system to more comprehensively understand and meet user needs. [30] Finally, although the DAIN model performs well on experimental datasets, its effectiveness in real-world applications still requires further validation and optimization. Future studies could consider testing the model on larger and more diverse real-world datasets to evaluate its generality and adaptability across different application scenarios. Additionally, as recommendation systems become widely applied across various industries, ensuring the protection of user privacy when handling sensitive data is an issue that cannot be overlooked. Future research could explore the integration of privacy protection mechanisms within the DAIN model, such as federated learning or differential privacy techniques, to ensure that while improving recommendation quality, user privacy and data security are maximized. [31] In summary, the DAIN model shows great potential and innovation in the field of personalized recommendation, but further optimization and exploration are needed in areas such as computational efficiency, robustness, multimodal data integration, and privacy protection in practical applications. These research directions will not only enhance the practical usability of the DAIN model but also provide new insights and technological support for the development of personalized recommendation systems. [32]

## 7. Conclusion

The Deep Adaptive Interest Network (DAIN) model offers a novel approach to personalized recommendation systems by integrating dynamic user interest modeling with context-aware learning. This model effectively captures user interest changes and provides personalized recommendations based on current context. Experimental results show that DAIN outperforms traditional algorithms like Matrix Factorization (MF), Neural Collaborative Filtering (NCF), and DeepFM, particularly excelling in key metrics such as NDCG@10 and HR@10.However, DAIN has limitations, including computational complexity, noise handling, multimodal data integration, and privacy concerns. Future research should focus on optimizing the model's efficiency, robustness, and integration of multimodal data while addressing privacy protection. Despite these challenges, the DAIN model shows great potential, and with further development, it can provide a solid foundation for more intelligent and secure recommendation systems.